\documentclass[english,twocolumn,superscriptaddress,showpacs]{revtex4}
\usepackage[T1]{fontenc}
\usepackage[latin9]{inputenc}
\usepackage{textcomp}
\usepackage{amsmath}

\makeatletter
\usepackage{changebar}
\usepackage{epsfig}
\usepackage{subfigure}
\usepackage{rotating}
\usepackage{amsfonts}

\psfigdriver{dvips}

\usepackage{babel}
\makeatother

\begin{document}

\title{Aging in Dense Colloids as Diffusion in the Logarithm of Time}
\author{Stefan Boettcher}
\email[]{www.physics.emory.edu/faculty/boettcher/}
\affiliation{Department of Physics, Emory University, Atlanta, GA, USA}
\author{Paolo Sibani}
\email[]{paolo.sibani@ifk.sdu.dk}
\affiliation{Institut for Fysik og Kemi, SDU, DK5230 Odense M, Denmark}

\begin{abstract}
The far-from-equilibrium dynamics of glassy systems share important
phenomenological traits.  A transition is generally observed
from a time-homogeneous dynamical regime to an aging regime where
physical changes occur intermittently and, on average, at a decreasing
rate. It has been suggested that a global change of the independent
time variable to its \emph{logarithm} may render the aging dynamics
homogeneous: for colloids, this entails diffusion
but on a logarithmic time scale.  Our novel analysis of experimental
colloid data confirms that the mean square displacement grows linearly
in time at low densities and shows that it grows linearly in the
logarithm of time at high densities. Correspondingly, pairs of
particles initially in close contact survive as pairs with a
probability which decays exponentially in either time or its
logarithm. The form of the Probability Density Function of the
displacements shows that long-ranged spatial correlations are very
long-lived in dense colloids.  A phenomenological stochastic model is
then introduced which relies on the growth and collapse of strongly
correlated clusters (``dynamic heterogeneity''), and which reproduces
the full spectrum of observed colloidal behaviors depending on the
form assumed for the probability that a cluster collapses
during a Monte Carlo update. In the limit where large clusters
dominate, the collapse rate is $\propto1/t$, implying a homogeneous,
$\log$-Poissonian process that qualitatively reproduces the
experimental results for dense colloids.  Finally an analytical
toy-model is discussed to elucidate the strong dependence of the
simulation results on the integrability (or lack thereof) of the
cluster collapse probability function.
\end{abstract}

\maketitle

\section{Introduction} 
Aging in amorphous materials has attracted widespread experimental,
simulational and theoretical interest for more than thirty
years~\cite{Struik78,Nordblad86,Rieger93,Kob00,Crisanti04,Sibani05}. As
a spontaneous off-equilibrium relaxation process, aging entails a
decrease of the free energy and, correspondingly, a slow change of
thermodynamic averages. E.~g., in numerical studies of models for
disordered magnets, the thermal energy decreases intermittently and,
on average, at a decelerating rate during the aging
process~\cite{Crisanti04,Sibani07,Sibani08,Christiansen08}.  Like
these low-temperature materials, colloidal suspensions at high density
(i.~e. high volume fraction) exhibit intermittent dynamics and a
gradual slowing down, here, in the rate at which particles move during
light scattering~\cite{Cipelletti00,Elmasri05} and particle tracking
experiments~\cite{Weeks00,Courtland03,Lynch08,Candelier09}. While the
phenomenology of colloidal aging is broadly similar to that of
thermally activated aging, spatially averaged quantities as energy and
particle density hardly change in colloids.  Furthermore, no external
field is required to elicit a measurable response.  Finally, the
motion of colloidal particles is time-homogeneous for sufficiently low
density, but for no accessible time scales and for no value of the
density does it look stationary or equilibrium-like.  Generalizations
of the Fluctuation Dissipation Theorem and the concept of effective
temperature\cite{Cugliandolo97,Castillo03}, developed to describe
thermally controlled aging in disordered magnets, are not applicable
to colloids, since the dichotomy between equilibrium-like fluctuations
and off-equilibrium dynamics is absent.  Reconciling differences and
similarities between aging in colloids and other glassy systems thus
requires a novel and broader approach .

\begin{figure*}
\hspace{-.85cm}
$\begin{array}{ccc}
\includegraphics[width=0.325\linewidth,height=0.26\linewidth]{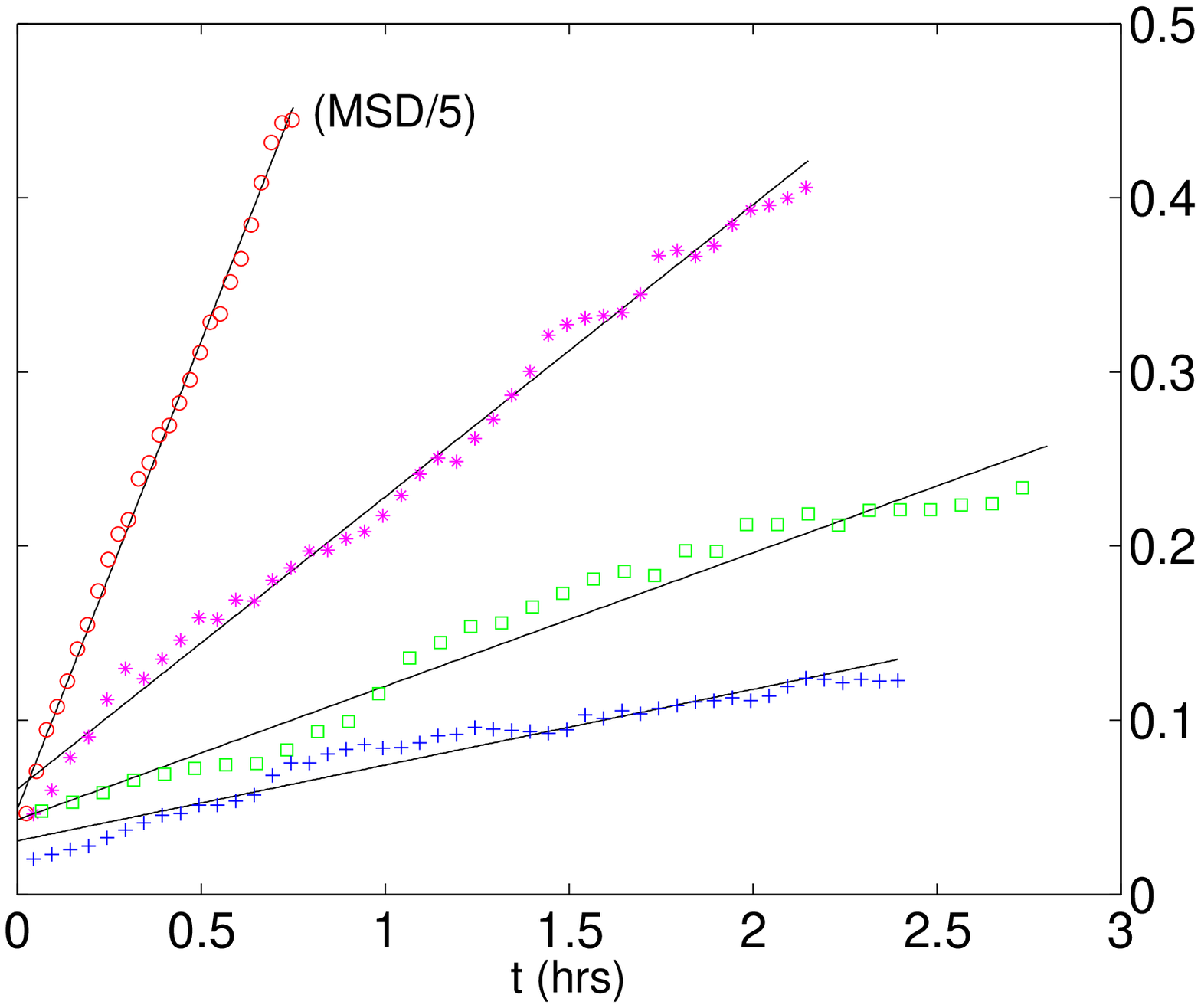}  &
\includegraphics[width=0.325\linewidth,height=0.26\linewidth]{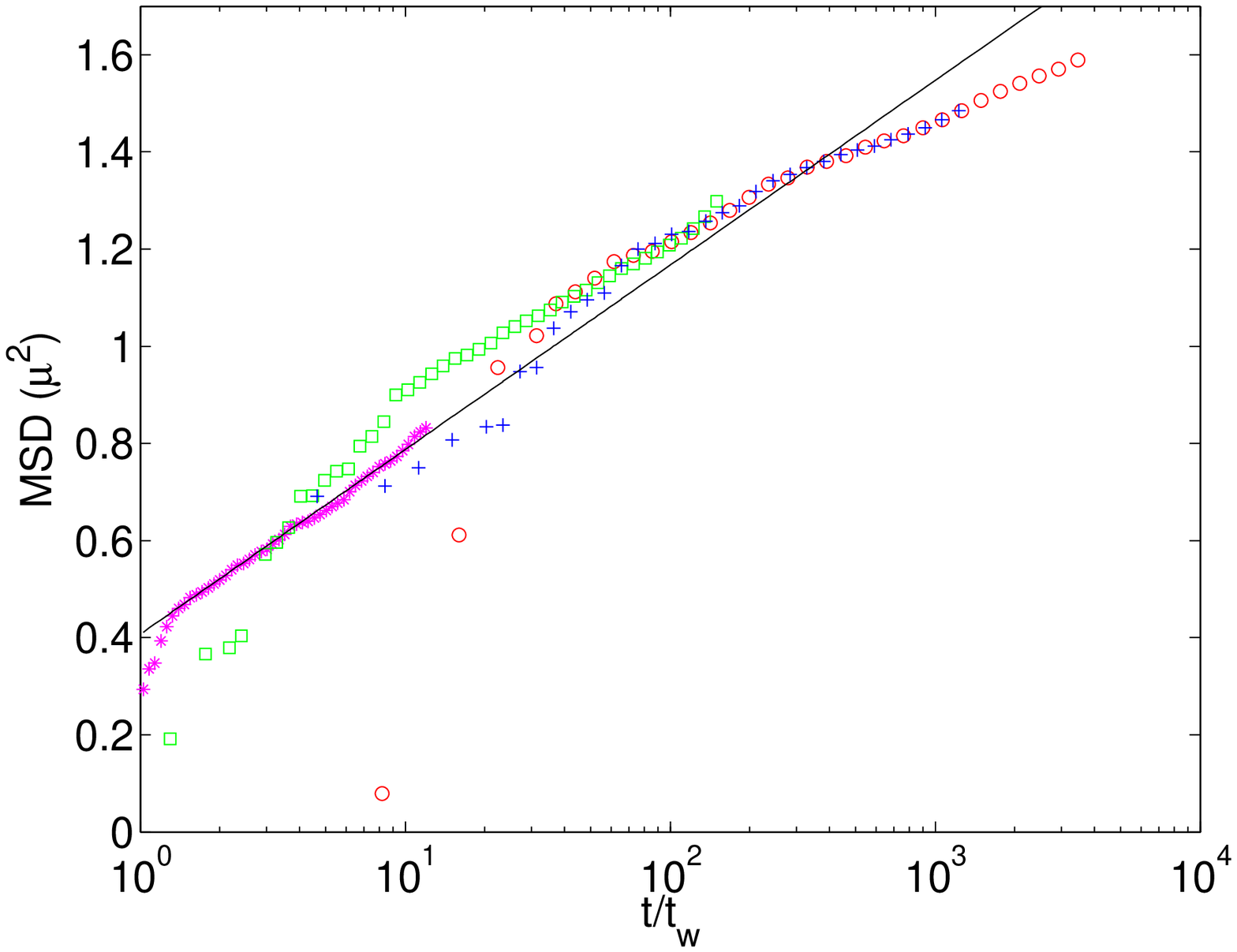} &
\includegraphics[width=0.325\linewidth,height=0.26\linewidth]{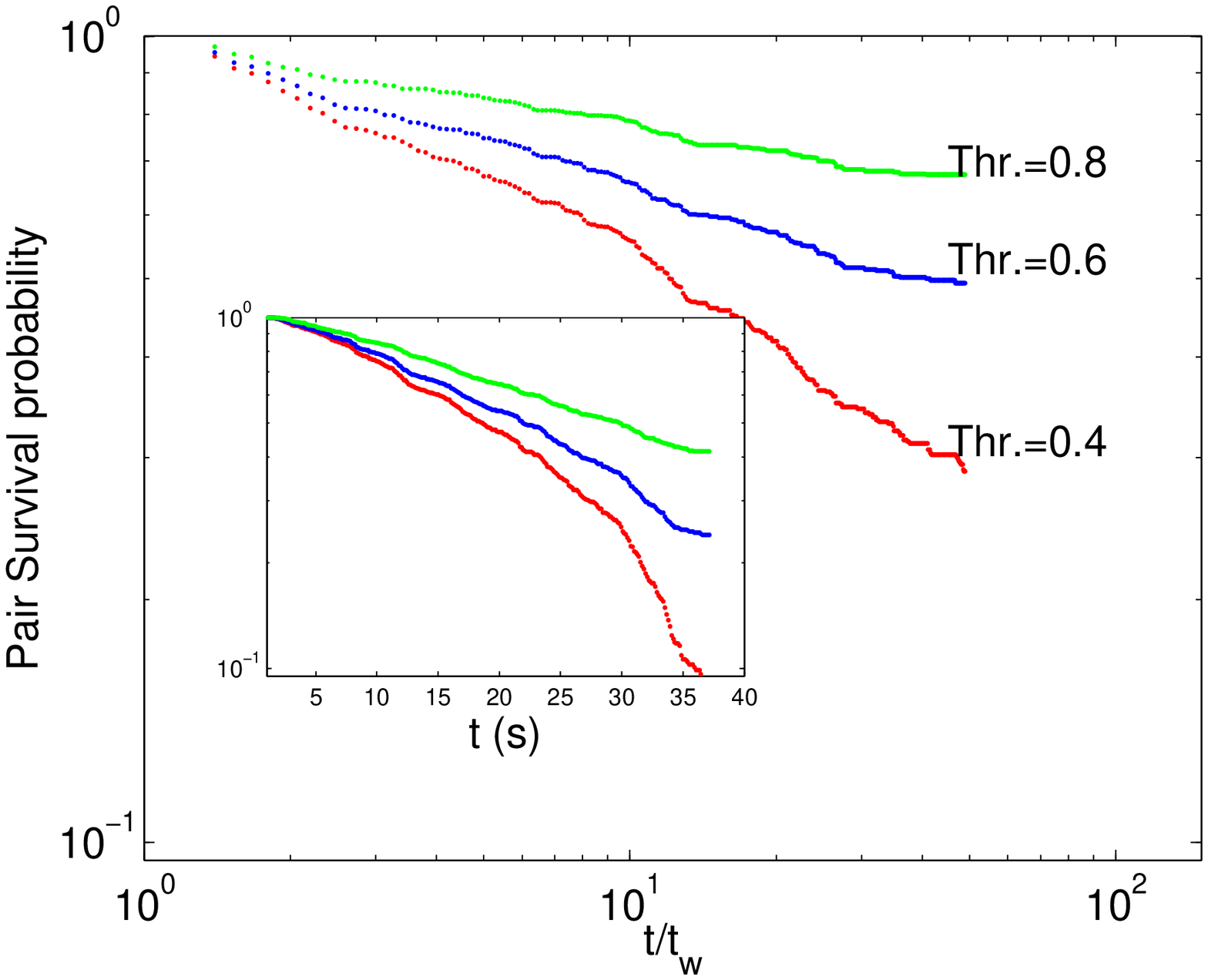} 
\end{array}
$
\caption{(Color online)
Left: Mean Square Displacement (in $\mu^2$) for colloids at low
density, volume fractions $\rho\approx0.46, 0.52, 0.53,0.56$, top to
bottom, all plotted versus time, in seconds. The top-most data set
has, for graphical reasons, been scaled down by a factor of five.
Middle: same quantity for dense colloids, where for all data
$\rho\approx0.62$.  Waiting times, in seconds: $t_w\approx 1,10,20,$
and $150$ for pluses, squares, circles, and stars, respectively.  Right:
At high density, ($\rho\approx 0.62$) the persistence decays as a
power law dependent on a separation-threshold $\theta$. The inset
shows the exponential decay at low density, $\rho\approx0.46$, for the
same thresholds.
}
\label{exp_stat} 
\end{figure*} 

In this work, the colloidal particle tracking experiments of Courtland
et al.~\cite{Courtland03} are re-analyzed, leading to the conclusion
that particle motion is diffusive in either time or logarithmic time,
at low or high densities, respectively. Dynamics properties, such as
the Mean Squared Displacement (MSD) and the persistence of tracer
particles, appear homogeneous when viewed as a function of an
appropriately re-scaled time variable. As the MSD, for instance, is a
sum of independent displacements, the correct choice of time variable,
hence, \emph{trivializes} colloidal dynamics\footnote{Of course, this
  observation does not dispute the profound consequences, such as
  memory effects, that can arise from the breaking of
  time-translational invariance.}.  To explain why that is,
a novel stochastic model is introduced with a single free parameter
controlling the transition from diffusive to log-diffusive
behavior. In the log-diffusive regime, the dynamics is driven by
increasingly rare events, called ``quakes'', whose temporal
distribution is homogeneous in logarithmic time. This links the
dynamical behavior of dense colloids to the notion of ``record
dynamics''\cite{Sibani03,Anderson04}, a relaxation paradigm already
applied to a number of complex
systems\cite{Oliveira05,Sibani05,Sibani06a,Christiansen08}.
 
Our experimental data analysis considers the MSD and the persistence,
i.~e.~the probability that a pair of particles initially close to each
other remain close for at least time $t$, under a number of different
conditions. The Probability Density Function (PDF) of the particle
displacement is likewise measured to illustrate intermittency and the
long range of the spatial correlations. The model encodes the spatial
heterogeneity in particle mobility, which is ubiquitous in glassy
systems~\cite{Kegel00,Weeks00,Mayer04,Candelier09}, in terms of the
formation and evolution of clusters of strongly correlated spatially
contiguous particles.  These clusters signal emerging ``dynamic
heterogeneities'' (DH), i.~e. a broad distribution of time-scales for
the survival and break-up of correlation patterns in different
regions~\cite{Weeks00}. The stochastic model abstracts the underlying
dynamics solely into a cluster survival probability: complex behavior
replaces simple diffusive behavior once clusters attain non-zero
probability to survive indefinitely under the update rule.  When the
probability per update to break up falls off exponentially with
cluster size, events in the model become well-separated in time, with
a log-Poissonian statistic. This reproduces the qualitative features
of the experimental data: the dynamics is intermittent and homogeneous
in logarithmic time.

\begin{figure*}
$
\begin{array}{cc}
\includegraphics[width=0.45\linewidth,height=0.35\linewidth]{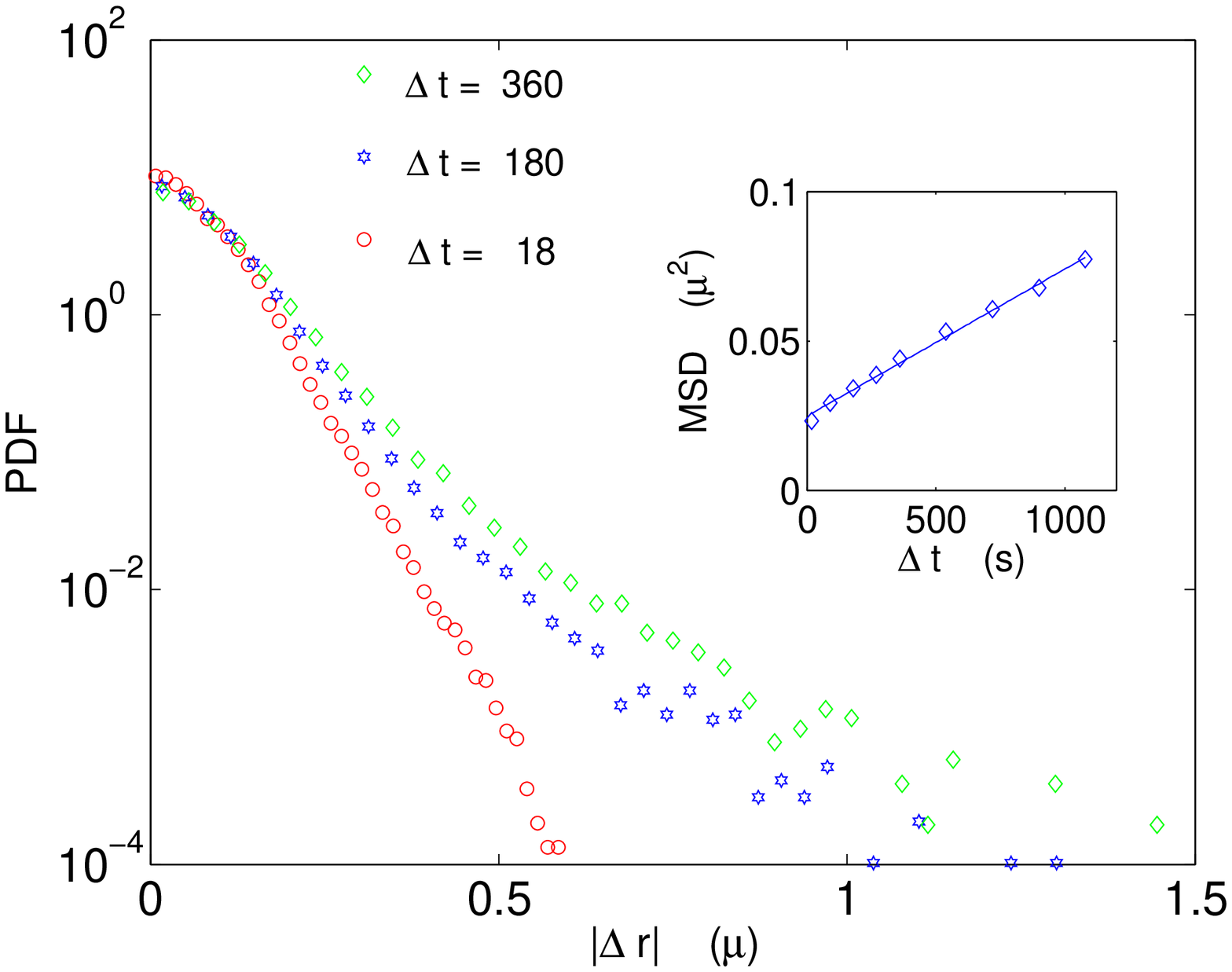}  &
\includegraphics[width=0.45\linewidth,height=0.35\linewidth]{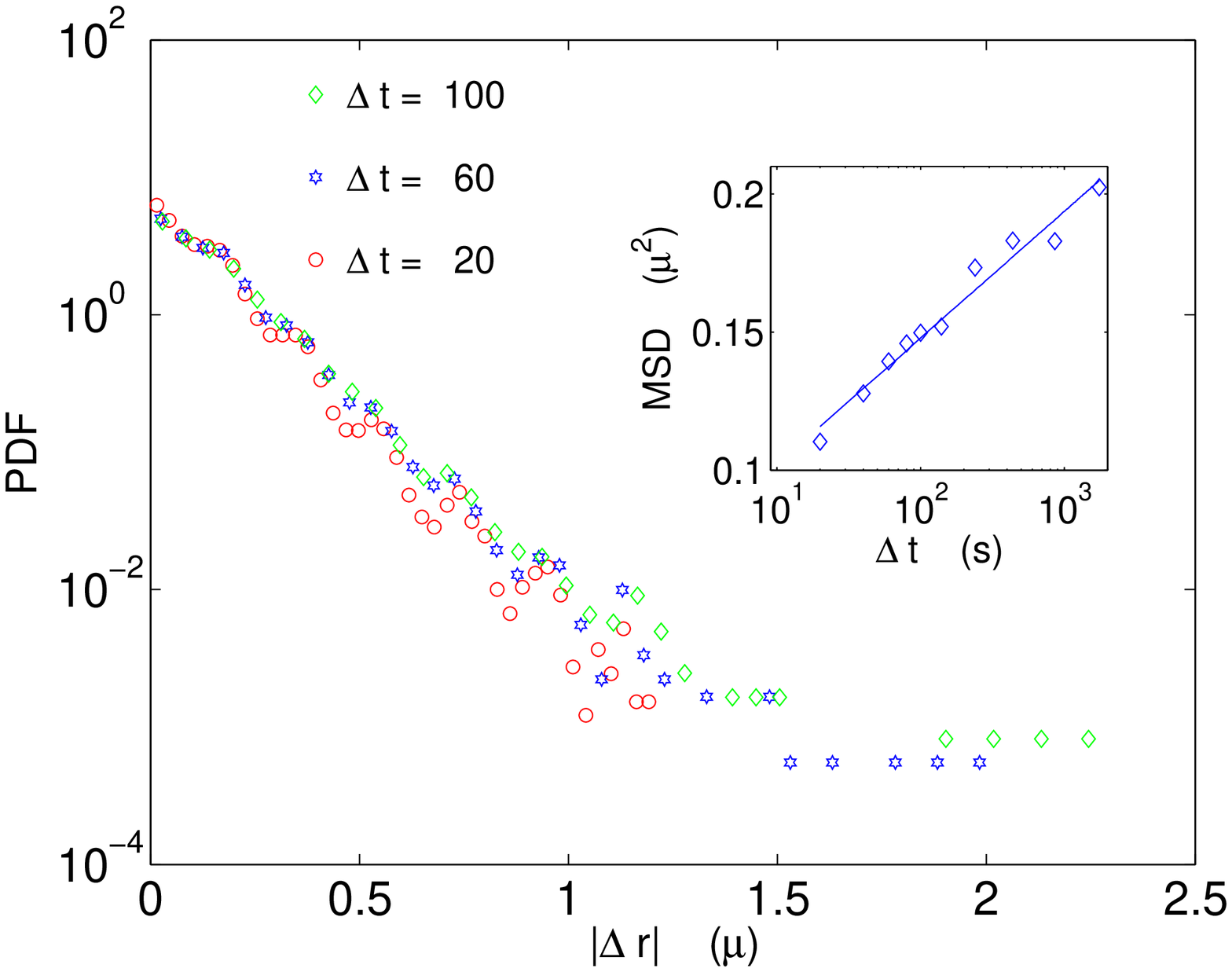}  
\end{array}
$
\caption{ (Color online)
The left panel pertains to the liquid, and the right one to the glassy
regime of a colloid.  Both panels display on a log-scale the PDF for
the absolute value $|\Delta r|$ of the particle displacement (in any
direction).  Three data sets are shown, corresponding to different
durations (in seconds) of the time interval $\Delta t$ over which the
displacement is measured. In the inserts, the Mean Square Displacement
(MSD) over a larger set of time intervals is plotted versus the
interval length, using a linear (left panel) and logarithmic (right
panel) abscissa.}
\label{intermittency} 
\end{figure*}  

\section{Experimental Data Analysis} 
Obtained using particle tracking techniques, the colloidal data of
Courtland et~al.~\cite{Courtland03} comprise the $3d$-trajectories of
a set of tagged colloidal particles.  The particle radius is $1.18$
microns, and the position coordinates are given in microns, a length
unit nearly equal to the particle radius.  Colloids were initially
centrifuged to obtain the desired density, and then briefly stirred.
The end of the stirring phase is considered as the origin of the time
axis, i.e., $t=0$. Without time-translational invariance, this choice
impacts the description of dense colloids.

The MSD, relative to any overall drift in the system, is calculated
from the trajectories as a function of time.  Fig.~\ref{exp_stat},
leftmost panel, shows the well-known diffusive behavior of low density
colloids.  The diffusion coefficient decreases monotonically with
increasing volume fraction.  MSD pertaining to high-density colloids
are plotted in the middle panel on a log horizontal scale versus the
scaling variable $t/t_w$, where $t_w, t_w \le t$ is the age at which
data collection commences.  Although a more precise knowledge of $t_w$
would facilitate a better data collapse, it is apparent that the
particles in a dense colloid clearly diffuse in \emph{logarithmic}
time.
  
The rightmost panel of Fig.~\ref{exp_stat} displays data for the
persistence defined above.  To ensure statistical independence, we
partition the system in $20\!\times\!20\!\times\!5$ identical
sub-volumes. At time $t_w$, a particle centrally located in each
sub-volume and its closest neighbor are tagged as a pair. If at time
$t>t_w$ the distance between the elements of a pair increases beyond a
given threshold value $\theta$ \emph{and} subsequently remains above
$\theta$, the pair is deemed to have broken up at $t$.  In the plot,
three $\theta$-values are considered.  Measured in units of particle
radius, these $\theta$ are large compared to the typical range of the
relative motion. The persistence decays as a power of $t$ for dense
colloids, i.~e.~exponentially in $\log(t)$, while at low density it
relaxes exponentially, as shown in the insert.

Finally, Fig.~\ref{intermittency} shows the empirical PDF (also known
as the self-part of the Van Hove distribution function) of the
absolute value of the particle displacement (in any direction) in the
liquid (left panel) and glassy (right panel) regimes of the
colloids. The vertical scale is logarithmic.  The three data sets
displayed correspond to different values of the time interval $\Delta
t$ over which the displacement is observed.  The inserts include more
values of $\Delta t$, and merely confirm the diffusive, respectively
log-diffusive, character of the dynamics previously discussed. Note
that the MSD does not cross the origin.  The particles already have
acquired some dispersion after the first sweep of the confocal
microscope, probably due to a residual stirring motion.

In similar analyses performed on a number of different
systems~\cite{Stariolo06,Chaudhuri07}, a central Gaussian part is
generally flanked by an exponential `non-Fickian' tail.  The left-most
part of the PDF should, by analogy, have a parabolic shape, and some
curvature is indeed visible in the left panel at small values of the
abscissa. However, the exponential `tail' extends right up to $|\Delta
r|= 0$ in the glassy regime.  Thus, long-ranged spatial correlations
survive over long times in dense colloids.  This very feature is
incorporated as a central hypothesis in the model introduced next.

\begin{figure*}
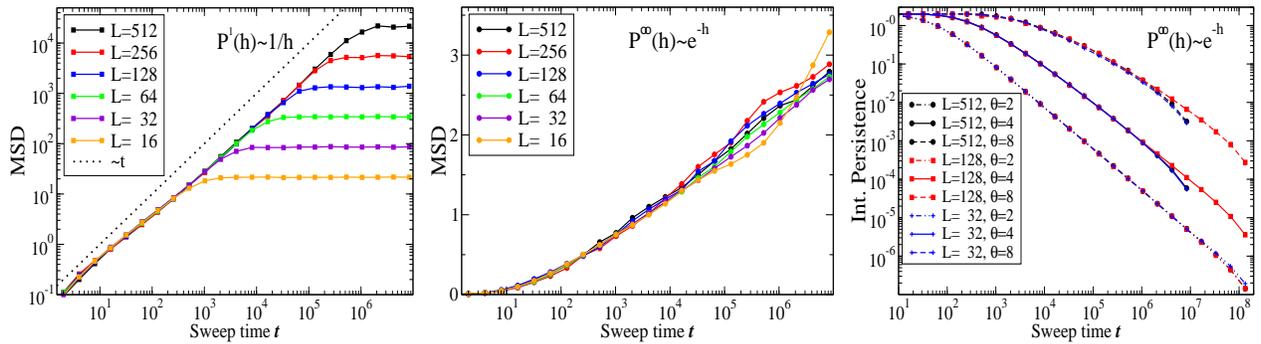

$\begin{array}{ccc}
\includegraphics[width=0.3\linewidth,height=0.25\linewidth]{MSD_new_hy1.eps}  &
\includegraphics[width=0.3\linewidth,height=0.25\linewidth]{MSD_new_exp.eps}  &
\includegraphics[width=0.3\linewidth,height=0.25\linewidth]{Persistence_Int_exp.eps}
\end{array}
$
\caption{(Color online)
Simulation results for different system sizes $L$ for $P^{1}(h) \sim
1/h$ (left) and $P^{\infty}(h) \sim \exp(-h)$ (center and right), to
be compared with Fig.~\ref{exp_stat}. For $\alpha=1$ (left) the MSD
behaves diffusively, represented by the dotted line of unit slope,
while for $\alpha=\infty$ the MSD grows linearly in logarithmic time,
with no sign of saturation. The persistence for $\alpha=\infty$
(right) exhibits a power-law, independent of the threshold $\theta$.}
\label{mod_stat} 
\end{figure*}  

\begin{figure*}
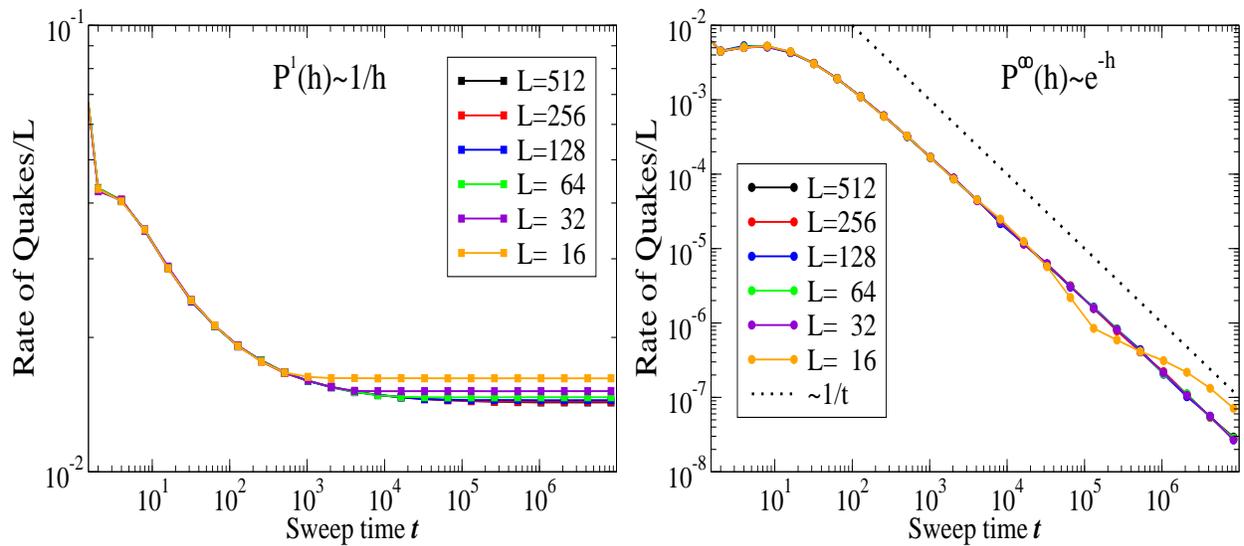

$\begin{array}{cc}
\includegraphics[width=0.45\linewidth,height=0.4\linewidth]{quakes_new_hy1.eps}  &
\includegraphics[width=0.45\linewidth,height=0.4\linewidth]{quakes_new_exp.eps} 
\end{array}
$
\caption{ (Color online)
Plot of the rate of cluster collapse vs. system age $t$ for $P^{1}(h)$
(left) and $P^{\infty}(h)$ (right). }
\label{rates} 
\end{figure*}  

\section{Cluster Model} 
The cluster model discussed below assumes that strong kinematic
constraints bind the particles together inside clusters that form a
spatially heterogeneous patchwork of particle properties.  The growth
and destruction of these clusters are the only mechanisms allowed for
the net displacements of particles.  The update dynamics is a Markov
chain where a randomly selected cluster of (integer) size $h$ either
survives intact or is destroyed with probability $P(h)$. By
assumption, $0\le P(h)\le 1$ decreases with $h$, i.~e. larger clusters
are more stable than smaller ones.  The particles released by a
collapsed cluster join neighboring clusters, and are activated to move
in real space by a unit step in a random direction.  Model variants
are characterized by different forms of $P(h)$.  Merely the
\emph{integrability} of $P(h)$ in $h$ discriminates between the
behaviors similar to those of low and high density colloids, and
without loss of generality we can limit ourselves to
\begin{equation}
P^{\alpha}(h)= \frac{1}{\sum_{k=0}^\alpha \frac{h^k}{k!}}, \quad \alpha=1,2 \ldots,
\label{decay_prob}
\end{equation}
an $\alpha$-family of models with $P^{\alpha=\infty}(h) = \exp(-h)$.

In our Monte Carlo simulations the clusters are arranged in a periodic
lattice of length $L$.  Initially, each lattice site holds uniformly a
cluster of size $h_i=\bar{h}$. The average number of particles bound
in clusters $\sum_{i=1}^L h_i/L={\bar h}$ remains conserved. Note that
these clusters do not per-se represent individual colloidal particles
and their motion, since particles typically occupy space uniformly,
with density fluctuations difficult to discern. Instead, ranges in
cluster sizes represent DH in particle mobilities, which can shift
dramatically even though actual particles hardly move.  Here,
heterogeneities arise \emph{dynamically} as larger clusters, once
formed, are more long-lived than smaller ones, and the particles which
belong to them correspondingly are less mobile.

We update $t\to t+1$  in parallel at all even and odd sites in alternating
order. Each site $i$ with $h_i>0$ is updated by drawing a univariate
random number $r_i$. If $r_i<P^\alpha(h_i)$, the cluster on that site
breaks up into two randomly split fractions that are respectively
added to the neighboring sites $i\pm1$, leaving $i$ empty.  To
simulate the actual (activated) motion of particles, we add (mutually
non-interacting) tracer particles which reside on lattice sites. Those
walk randomly $i\to i\pm1$ \emph{iff} there is a cluster on their site
$i$ that happens to break up. In particular, a tracer particle can be
stranded on a cluster-free site for a long time until a nearby cluster
breaks up such that its scattered ``debris'' can reactivate that
particle.

{}Fig.~\ref{mod_stat} shows the MSD averaged over all tracer particles
for $\alpha=1$ and $\alpha=\infty$.  The former behaves diffusively
(up to a cutoff scale $t_{\rm st} \propto L^2$). In the latter, no
stationary state is approached and the behavior is diffusive in
\emph{logarithmic time}, as in the corresponding experimental result
shown in Fig.~\ref{exp_stat}. Fig.~\ref{mod_stat} also depicts
persistence curves, i.~e. probabilities that pairs of tracer particles
initially located at the same grid point remain within a separation of
$\theta$ sites after $t$ sweeps. As in the experiments, the
(asymptotic) decay goes from being exponential in time for $\alpha=1$
(not shown~\cite{BoSiLong}) to being exponential in logarithmic time
when the cluster collapse probability goes from hyperbolic to
exponential at $\alpha=\infty$, independent of $\theta$. Unlike the
experiments, changing $\theta$ does not effect the power-law exponent
but modifies the time for its onset.

\section{Record Dynamics}
As the collapse of large clusters controls all aspects of the model in
the aging regime, the statistics of such \emph{quake}-events is key to
the analysis, a situation \emph{identical} to aging in quenched
glasses~\cite{Sibani07}. In turn, the fast collapse of small clusters
provides a background of mobile particles diffusing within regions
(``cages'') bounded by large clusters. The divide between small and
large clusters was empirically taken to be $\max({\bar h},\sigma(t))$,
where $\sigma(t)$ is the (growing) standard deviation of the cluster
size distribution.  The quake rate versus time, scaled by the system
size $L$, is shown in Fig.~\ref{rates} for our simulations.  For
$P^1(h)$ the rate approaches a constant on a time scale independent of
system size. For $P^\infty(h)$, the stationary regime is in fact
invisible even for the smallest systems, and the dynamics is thus
purely non-stationary with a decelerating quake rate decaying as
$1/t$.  Assuming that successive quakes are statistically independent
events (demonstrated elsewhere~\cite{BoSiLong}), they provide the true
clock of the dynamics, and the particle MSD simply becomes
proportional to the time \emph{integral} of the quake rate: linear in
time in the stationary (liquid) regime, and $\sim\log(t)$ in the aging
regime. Clearly then, the MSD in the aging regime between times $t_w$ and $t$
scales as $t/t_w$, a so-called full aging behavior
 which is consistent with the experimental findings.

The model dynamics for the exponential collapse probability can be
linked to \emph{record
  dynamics}~\cite{Sibani93a,Sibani03,Sibani04a,Anderson04,Oliveira05,Sibani06a,Christiansen08},
using the wide separation of time scales associated with the collapse
of clusters of different sizes. Let us first note that the empirically
estimated boundary between small (i.e. irrelevant) and large
(i.e. important) clusters, $h(t) \approx \ln t$ can in this case also
be obtained by equating to the time $t$ the  number of queries that a cluster of height
$h$ on average  survives. The pace of the dynamics is
always controlled by the smallest among the large clusters,
henceforth the SL cluster. The
latter is the first one to collapses with overwhelmingly high
probability.  Let the corresponding quake happen on a time scale
$t^{(0)}$.  The new SL cluster  
can either be pre-existing or can be formed through the accretion
process of the debris produced by the last quake, thus
recapitulating all previous history, an accretion process which is now
instantaneous on the time scale $t^{(0)}$. Since the cluster size
distribution is essentially continuous, the new SL cluster will
only be slightly larger than its predecessor, a hallmark of ``marginal
stability increase''~\cite{Sibani08}.  Hence, a record-sized random
number (corresponding to a record sized re-arrangement in the colloid)
will suffice to destroy the new SL cluster.  At time-step $t$
the probability is $1/t$ that such event will happen.  Statistical
independence is assured by the wide separation of break-up
events. Thus, inter-event times $\tau^{(n)} =\ln t^{(n)}\!-\!\ln
t^{(n-1)}\!=\!\log\left[t^{(n)}/t^{(n-1)}\right]$ become Poissonian
distributed~\cite{Sibani93a,Krug05}.  To ascertain the log-Poisson
nature of the quake time distribution, we collect the quake times
$t^{(n)}$ and show (elsewhere~\cite{BoSiLong}) that the series
$(\tau^{(1)},\tau^{(2)},\ldots, \tau^{(n)}, \ldots)$ of
\emph{logarithmic waiting times} $\tau^{(n)}$ is $iid$ with the same
exponential distribution.  In contrast, when the cluster collapse
probability becomes long-tailed, $\alpha<\infty$, break-up events
overlap and long-lived clusters are deemphasized.

\section{Mean-Field Description}
Finally, we provide a mean-field model of the  clustering dynamics that elucidates the
preceding results.  The fundamental quantity considered is the average
number of clusters $n_{k,t}$ of \emph{binary} size $h=2^{k}$
($k=0,1,2,\ldots$) extant at time $t$.  For $k>0$, those $k$-clusters
are immobile and break up with probability $p_k\sim P(2^k)$, but
merely into two $k-1$-clusters. Eq.~(\ref{decay_prob}) reduces to
$p^\alpha_k\sim\left(2^k\right)^{-\alpha }$.  Integrability of $p_k$
is conferred via $P(h)dh\sim p_kd(2^k)\propto2^kp_kdk$,
i.~e. $\sum_k2^kp^\alpha_k<\infty$ for $\alpha>1$. To write down a
self-consistent equation for $n_{k,t}$, we stipulate that during each
\emph{sweep} the number of $k$-clusters $n_{k,t}$ for all $k>0$
decreases through break-ups by a factor $1-p_k$, yet, it also
increases through the break-up of $k+1$-clusters into \emph{two}
$k$-clusters (with weight $2p_{k+1}$). Unit ($k=0$)-clusters are
indivisible but mobile. Larger $k$-clusters further grow via accretion
of those small mobile clusters, but (somewhat artificially) only in
groups of $2^{k}$ with probability $\propto1/2^k$. During a sweep,
\emph{all} unit clusters attach somewhere, i.~e. $n_{0,t+1}$ is
entirely unrelated to $n_{0,t}$. Hence, we obtain
\begin{eqnarray} 
n_{k,t+1} & = & n_{k,t}\left(1-p_k\right)+2p_{k+1}
n_{k+1,t}+\frac{n_{k-1,t}n_{0,t}}{2^{k}N_t}\qquad~
\label{eq:DH}
\end{eqnarray}
for $k>0$, and $n_{0,t+1} = 2p_1 n_{1,t}$. The last term describes
growth through the accretion of all unit clusters $n_{0,t}$, equally
shared between all clusters, $N_t=\sum_{k'=0}^\infty n_{k',t}$. Its
form ensures the conservation of $\sum_{k'=0}^\infty
2^{k'}n_{k',t}(={\bar h}L)$, as used in our simulations above.

Defining $a_{k,t}=2^kp_kn_{k,t}$, we rewrite Eq.~(\ref{eq:DH}) as
\begin{eqnarray} 
\frac{1}{p_k}\left(a_{k,t+1}-a_{k,t}\right) & = &
a_{k+1,t}-a_{k,t}+a_{k-1,t}{\cal K}_{k-1}{\cal T}_t.\qquad~
\label{eq:DHa}
\end{eqnarray}
Amazingly, our analysis finds that one of the factors, ${\cal K}_{k} =
1/\left(2^kp_k\right)$ or ${\cal T}_t=n_{0,t}/N_t$, is \emph{always}
sufficiently small asymptotically as to ignore the last term of
Eq.~(\ref{eq:DHa}). The remaining linear equation has stationary
solutions (for $t\sim t+1\to\infty$) entailing $a_{k+1}\sim a_k\sim
const$, i.~e. $n_k\sim 1/\left(2^kp_k\right)$. Since the stationary
cluster size distribution must have a finite integral (after all,
$N_t\leq L<\infty$), $n_k$ must decrease sufficiently, demanding
$p_k\gg2^{-k}$ or $\alpha<1$. In this regime, ${\cal T}_t\to{\cal
  T}_\infty\leq1$ but $a_{k-1}{\cal K}_{k-1}\ll a_k$ justifies
dismissal of the last term in Eq.~(\ref{eq:DHa}). Conversely, for
$2^kp_k\ll1$ ($\alpha>1$), the aging regime, numerics suggests that
$n_{k,t}$ is highly localized at some increasing $k=k(t)$: clusters
grow and attain a typical size $h(t)$. Unit-size clusters hardly ever
occur, $n_{0,t}=0$ almost always, and although ${\cal K}_k$ grows
exponentially, ${\cal T}_t\approx0$ annihilates the last
term. Rescaling $\tau=p_kt$ then turns Eq.~(\ref{eq:DHa}) into a
``wave'' equation,
$\left[\partial_\tau-\partial_k\right]a(k,\tau)\approx0$, with
characteristic $k\sim\tau=p_kt$. Using $p_k=p^\alpha_k$ and
$h\sim2^k$, we extract the dominant growth-law for cluster sizes,
$h(t)\sim t^{1/\alpha}$, or for the event-rates (quakes) of
cluster-size increases:
\begin{equation}
\frac{\partial h}{\partial t}\sim t^{\frac{1}{\alpha}-1},\qquad(\alpha>1).
\label{eq:quakerate}
\end{equation}
The scaling reproduces for $\alpha\to1^+$ and for $\alpha\to\infty$
the simulation results in Fig.~\ref{rates} (and, correspondingly,
those for any $\alpha$ in-between~\cite{BoSiLong}).  While the
equations do not describe the displacement of tracer particles
explicitly, the discussion above implies that the quake rate sets the
scale for spread in time for \emph{any} observable, be it $h(t)$ or
MSD: logarithmic for $\alpha\to\infty$ and linear (with a cut-off) for
$\alpha\leq1$.

\section{Conclusions}
Experimental data are  analyzed showing that colloidal  motion
is diffusive in either time or logarithmic time, depending on
colloidal density. A model explaining these findings is presented,  based on 
the survival probability of clusters, i.~e.   highly correlated
groups of particles, representing  dynamically  heterogeneous 
regions of different mobility: a net displacement  of a particle 
 in the model requires the collapse of the cluster to which 
the particle  belongs, i.~e. a re-arrangement
of   (possibly large and increasing) correlation patterns. 
The survival probability of a cluster increases with its size.
Once large clusters obtain the capacity to survive indefinitely,
$P(h)\ll1/h$ for $h\to\infty$, non-stationary, aging
behavior arises, characterized by intermittency and memory
effects. 
Remarkably, the $t/t_w$ scaling of the 
experimental data (full aging) is  only 
achieved  in the model in the limit of an exponential $P(h)$, corresponding
to the extreme value $\alpha = \infty$ of its parameter. In that
sense, tuning a parameter such as $\alpha$, say, by changing the
volume fraction $\rho$ in a colloid, does not describe a sharp phase
transition. But, presumably, the transition between both extremes
occurs over a nearly unobservable interval of $\rho$.

{~}

{\bf Acknowledgments:}  
We thank Eric Weeks for helpful discussion and access to his data.  SB
is supported by the U.~S.~National Science Foundation through grant
DMR-0812204. PS thanks the Physics department of Emory University
for its hospitality and its Emerson Center for financial support.
\bibliographystyle{unsrt}
\bibliography{SD-meld}

\end{document}